\documentclass[11pt]{article}
\newtheorem{proposition}{Proposition}

\newenvironment{thmproof}[1]
{\noindent\hspace{2em}{\it #1 }}

\usepackage{epsf,cite,amsmath,amscd,amssymb,graphics,latexsym,multicol}
\usepackage{color}

\begin{document}
\bibliographystyle{ieeetr}
\renewcommand{\textfraction}{0}
\renewcommand{\baselinestretch}{1}
\title{\huge  \textbf{A Unified Power Control Algorithm for\\ Multiuser Detectors in Large Systems:\\ Convergence and Performance}\footnote{This research was
supported by the National Science Foundation under Grant
ANI-03-38807.}}
\author{\normalsize Farhad Meshkati, H. Vincent Poor and Stuart C. Schwartz \\
\small Department of Electrical Engineering \\ \small Princeton
University \\
\small Princeton, NJ 08544 USA\\
\small \{meshkati,poor,stuart\}@princeton.edu \vspace{0.2cm}\and
\normalsize Dongning Guo \\
\small Department of Electrical Engineering and Computer Science \\
\small Northwestern University \\
\small Evanston, IL 60208 USA\\
\small dguo@northwestern.edu \vspace{-0.3cm}}

\date{}
\maketitle \thispagestyle{empty} {\begin{abstract}\small{A unified
approach to energy-efficient power control, applicable to a large
family of receivers including the matched filter, the
decorrelator, the (linear) minimum-mean-square-error detector
(MMSE), and the individually and jointly optimal multiuser
detectors, has recently been proposed for
code-division-multiple-access (CDMA) networks. This unified power
control (UPC) algorithm exploits the linear relationship that has
been shown to exist between the transmit power and the output
signal-to-interference-plus-noise ratio (SIR) in large systems.
Based on this principle and by computing the multiuser efficiency,
the UPC algorithm updates the users' transmit powers in an
iterative way to achieve the desired target SIR. In this paper,
the convergence of the UPC algorithm is proved for the matched
filter, the decorrelator, and the MMSE detector. In addition, the
performance of the algorithm in finite-size systems is studied and
compared with that of existing power control schemes. The UPC
algorithm is particularly suitable for systems with randomly
generated long spreading sequences (i.e., sequences whose period
is longer than one symbol duration).}
\end{abstract}}
\normalsize

\section{Introduction}

Power control is used for interference management and resource
allocation in both the downlink and the uplink of
code-division-multiple-access (CDMA) networks. Power control for
CDMA systems has been studied extensively over the past decade
(see for example \cite{Foschini93, Bambos95, Yates95, Ulukus97,
HanlyTsePC, ShitzVerdu01, Andrews02}).  In the uplink, the purpose
of power control is for each user to transmit just enough power to
achieve the required quality of service (QoS) without causing
unnecessary interference.

Multiuser receivers are expected to be deployed in future wireless
systems, especially in the uplink, because of their superior
performance to the conventional matched filter \cite{VerduBook98}.
Because of this, power control for multiuser detectors has
attracted attention in recent years. In particular, power control
algorithms for the linear minimum-mean-square-error (MMSE)
detector and successive interference cancellation (SIC) receiver
have been proposed in \cite{Ulukus97} and \cite{Andrews02},
respectively. In the proposed schemes, the output
signal-to-interference-plus-noise ratio (SIR) is measured for each
user and then the user's transmit power is adjusted to achieve the
desired target SIR.

Almost all of the power control schemes proposed so far have a
specific receiver in mind. Reference \cite{Meshkati_SPWC},
however, proposes a unified power control (UPC) algorithm which is
applicable to a large family of multiuser detectors. This
algorithm is based on the large-system results in \cite{Guo05}
where a linear relationship between the input power and the output
SIR has been shown to exist for a family of multiuser detectors.
Members of this family include many well-known receivers such as
the matched filter (MF), the decorrelator (DE), and the MMSE
detector as well as the individually optimal (IO) and jointly
optimal (JO) multiuser detectors \cite{VerduBook98}. This linear
relationship, which is characterized by the multiuser efficiency,
is exploited in obtaining the proposed power control algorithm. In
\cite{Meshkati_SPWC}, the convergence of the UPC algorithm is not
proved and is only demonstrated through simulation. In this paper,
we prove the convergence of the UPC algorithm for the matched
filter, the decorrelator and the MMSE detector. In addition, we
study the performance of the UPC algorithm in finite-size systems
and compare it with that of the existing power control algorithms.
Since the UPC algorithm is based on large-system results, it does
not depend on instantaneous spreading sequences. Therefore, it is
particularly useful in CDMA systems with long spreading sequences
(e.g., the uplink of cdma2000). In systems with long spreading
sequences (i.e., sequences whose period is longer than one symbol
duration), other power control algorithms may need to adjust the
users' transmit powers symbol-by-symbol even when all the users'
channels stay fixed. The UPC algorithm, on the other hand, does
not need to update the powers if the channel gains stay the same.
Since the true SIR does depend on the spreading sequences, as the
spreading sequence changes from one symbol to the next, the SIR
achieved by the UPC algorithm deviates from the target SIR.
However, we show that if the processing gain is reasonably large,
the SIR achieved by the UPC algorithm stays close to the target
SIR most of the time.

The organization of this paper is as follows. In
Section~\ref{multiuser}, we provide the system model and some
relevant background on multiuser detection in large systems. In
Section~\ref{power control}, the unified power control algorithm
is presented and its convergence for linear receivers is proved.
The performance of the UPC algorithm in finite-size systems is
studied in Section~\ref{performance} and simulation results are
presented in Section~\ref{simulation}. Finally, conclusions are
given in Section~\ref{conclusions}.

\section{Multiuser Detection in Large Systems} \label{multiuser}

We consider the uplink of a synchronous DS-CDMA system with $K$
users and processing gain $N$. Let $p_k$, $h_k$, and $\gamma_k$
represent the transmit power, channel gain and output SIR,
respectively, for user $k$. Also, define
\begin{equation}
\Gamma_k = \frac{p_k h_k}{ \sigma^2}
\end{equation}
as the received signal-to-noise ratio (SNR) for user $k$ where
$\sigma^2$ here is the background noise power (including other
cell interference). The received signal (after chip-matched
filtering) sampled at the chip rate over one symbol duration can
be represented as
\begin{equation} \label{eqsys}
  {\mathbf{Y}} = \sum^K_{k=1} \sqrt{\Gamma_k} X_k {\mathbf{s}}_k +
  {\mathbf{W}}  , 
\end{equation}
where $\mathbf{s}_k$ and $X_k$ are the spreading sequence and
transmitted symbol of user $k$, respectively. We assume random
spreading sequences for all users, i.e., $ {\mathbf{s}}_k =
\frac{1}{\sqrt{N}}[v_1 ... v_N]^T$, where the $v_i$'s are
independent and identically distributed (i.i.d.) random variables
taking values \{$-1,+1$\} with equal probabilities. The $X_k$'s
are assumed to be i.i.d. with probability density $p_X$. In
(\ref{eqsys}), $\mathbf{W}$ is the normalized noise vector
consisting of independent standard Gaussian entries.

\begin{figure}
\begin{center}
\leavevmode \hbox{\epsfysize=3cm \epsfxsize=9cm
\epsffile{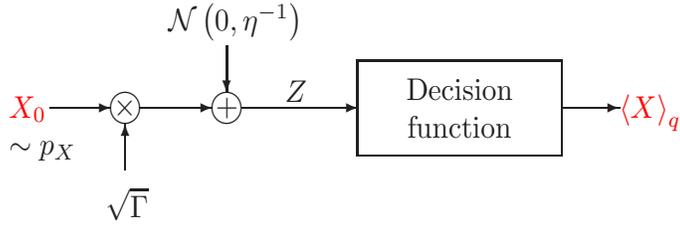}}
\end{center}
\vspace{-0.3cm}\caption{Equivalent single-user channel for the
multiuser CDMA channel.} \label{figsys2}
\end{figure}

It is shown in \cite{Guo05} that in large systems with random
spreading sequences, the multiuser channel combined with the
uplink detector can be decoupled into equivalent parallel
single-user Gaussian channels with some degradation in the SNR due
to multiple-access interference as shown in Fig.~\ref{figsys2}. By
a large system, we refer to the limit in which the number of users
and the spreading factor in a CDMA system both tend to infinity
but with a fixed ratio, i.e.,
\begin{equation}\label{eq2zz}
\lim_{K,N\rightarrow \infty} \frac{K}{N}= \alpha \ .
\end{equation}

The degradation parameter, known as the \emph{multiuser
efficiency}, completely characterizes the performance of each
individual user and can be determined by solving some fixed-point
equations, which we shall discuss shortly. This ``decoupling
principle" holds for a large family of detectors, called posterior
mean estimators (PMEs), which includes the matched filter, the
decorrelator, and the MMSE detector, as well as the individually
and jointly optimal multiuser detectors. This decoupling result
implies that in large systems there is a linear relationship
between a user's transmit power and its output SIR characterized
by the multiuser efficiency, $\eta_k$:
\begin{equation}
  \gamma_k = \eta_k \Gamma_k \ .  \label{eqSIR}
\end{equation}
This relationship is mainly due to the fact that in a large
system, as the user's transmit power changes, the interference
seen by that user essentially stays the same as long as the
overall distribution of the received powers remains the same. This
result generalizes to multirate and multicarrier systems as well
\cite{GuoMultirate, GuoMulticarrier}. In general, the multiuser
efficiency depends on the received SNRs, the spreading sequences
as well as the type of detector. However, in the asymptotic case
of large systems, the dependence on the spreading sequences
disappears and the received SNRs affect $\eta$ only through their
distribution\footnote{Since different users may have different
receiver types, the multiuser efficiency, in general, may be
different for different users.}. Note that although (\ref{eqSIR})
is true only in the large-system limit, it is a very good
approximation for most finite-size systems of practical interest.
In this paper, we focus on the implication of this result in
designing a unified power control scheme for multiuser detectors.
The linear dependence of the user's SIR on its transmit power can
also be used to generalize the results of \cite{Meshkati_Tcomm} to
nonlinear detectors such as the individually and jointly optimal
multiuser detectors (see \cite{Meshkati_SPWC}). In
\cite{Meshkati_Tcomm}, it has been shown that if we model power
control as a non-cooperative game with a utility function that
measures the energy efficiency of the network, then, for all
linear receivers, this game has a unique Nash equilibrium that is
SIR-balanced.

It is shown in \cite{Guo05} that for a particular class of PME,
which contains many popular receivers such as the conventional
matched filter, the decorrelator, and the MMSE detector as well as
the individually and jointly optimal detectors, the large-system
multiuser efficiency is obtained by solving the following joint
equations:
\begin{eqnarray}
  \eta^{-1} &=& 1 \ + \alpha \ {\mathbb{E}} \{\Gamma  \cdot {\mathcal{E}}(\Gamma;\eta,\xi)\} , \label{eqA} \\
  \xi^{-1} &=& \varrho^2 + \alpha \ {\mathbb{E}} \{\Gamma \cdot
  {\mathcal{V}}(\Gamma;\eta,\xi)\} \label{eqB},
\end{eqnarray}
where $\alpha=\lim_{K, N\rightarrow\infty}\frac{K}{N}$, and
$\varrho$ is a parameter determined by the receiver type. In
(\ref{eqA}) and (\ref{eqB}), $\mathcal{E}$ and $\mathcal{V}$ are
functions that can be easily computed given the distribution of
the transmitted symbols and the type of receiver; and the
expectations are taken with respect to the received SNR
distribution, $P_{\Gamma}$. For some popular detectors, the
multiuser efficiency is given as the solution of the following
equations (see \cite{Guo05}): {\small{\begin{eqnarray}
\eta^{MF} &=& \frac{1}{1+\alpha {\mathbb{E}}\{\Gamma\}} , \label{MF eta}\\
\eta^{DE} &=& 1-\alpha  \ \ \ \textrm{for} \ \ \alpha<1 ,
\label{DE eta} \\
 \frac{1}{\eta^{MMSE}}&=& 1 +\alpha {\mathbb{E}}\left\{
\frac{\Gamma}{1+\eta^{MMSE}\Gamma}\right\} , \label{MMSE eta}\\
\frac{1}{\eta^{IO}}&=& 1 +\alpha {\mathbb{E}}\left\{ \Gamma \left[
1- \int_{-\infty}^{+\infty} \frac{e^{-\frac{z^2}{2}}}{\sqrt{2\pi}}
\tanh \left(\eta^{IO}\Gamma - z
\sqrt{\eta^{IO}\Gamma}\right)\textrm{d}z\right]\right\} .
\label{IO eta}
\end{eqnarray}}}
Except for \eqref{IO eta}, which assumes binary inputs (i.e., BPSK
modulation), the rest are valid for all input distributions.

\section{The Unified Power Control Algorithm}\label{power control}

Recall from Section~\ref{multiuser} that in large systems, the
output SIR of user $k$ is given by
\begin{equation}
\gamma_k = \eta_k \Gamma_k= \eta_k  \frac{p_k h_k}{ \sigma^2} \ \
\textrm{for} \ \ k=1,\cdots,K .
\end{equation}

The objective of the UPC algorithm is for each user to iteratively
adjust its transmit power in order to reach an output SIR equal to
$\gamma^*$. While here we have assumed that the target SIR is the
same for all users, the UPC algorithm is general in the sense that
it can be applied to the case of unequal target SIRs. The
algorithm is also applicable to multirate systems. The description
of the UPC algorithm is as follows.\vspace{0.3cm}

\textbf{Algorithm 1}: The Unified Power Control (UPC) Algorithm:
\begin{enumerate}
\item \emph{n}=0, start with initial powers
$p_1(0),\cdots,p_K(0)$. \item Based on the power profile, compute
the multiuser efficiency, $\eta_k(n)$, using
\eqref{eqA}~and~\eqref{eqB}. \item Update the powers using
$p_k(n+1) = \frac{1}{\eta_k(n)} (\frac{\gamma^* \sigma^2}{h_k})$
for $k=1, ... , K$. \item \emph{n}=\emph{n}+1, stop if
convergence; otherwise, go to Step 2.
\end{enumerate}

In Step 2, while finding an analytical expression for $\eta_k$ is
difficult for most multiuser detectors, $\eta_k$ can be easily
obtained using numerical methods. Note that (\ref{eqA}) and
(\ref{eqB}) need to be solved only once per iteration for each
user. The uplink receiver (e.g., base station) can, for example,
compute $\eta_k$ and feed it back to the user terminal. Also, if
all the users have the same type of receiver, $\eta_k$ will be
independent of $k$ and, hence, we need to solve for $\eta$ only
once per iteration which greatly reduces the computational
complexity of the algorithm. The above algorithm is applicable to
a large family of receivers which includes many popular receivers
such as the matched filter, the decorrelator, and the MMSE
detector as well as individually and jointly optimal multiuser
detectors. In actual implementation of the algorithm, the
expectations in \eqref{eqA} and \eqref{eqB} can be replaced by
summations over all users' received SNRs (or their estimates). For
example, \eqref{MMSE eta} can be expressed as $\frac{1}{\eta} = 1
+ \frac{\alpha}{K} \sum^K_{k=1} ( \frac{\Gamma_k}{1+\eta \Gamma_k}
)$.

We now prove the convergence of the UPC algorithm for the matched
filter, the decorrelator, and the MMSE detector\footnote{The
convergence proof for a general receiver remains an open problem.
To prove the convergence in the general case, one has to deal
directly with \eqref{eqA} and \eqref{eqB} which are difficult to
work with.}. To prove the convergence, let
${\mathbf{\Gamma}}=\left[\Gamma_1, \cdots , \Gamma_K\right]$ and
let us define an interference function,
${\mathbf{I}}({\mathbf{\Gamma}})=\left[I_1({\mathbf{\Gamma}}) ,
\cdots , I_K({\mathbf{\Gamma}})\right]$, where
\begin{equation}
I_k({\mathbf{\Gamma}})=\frac{\gamma^*}{\eta_k({\mathbf{\Gamma}})}
\ .\vspace{-0.2cm}\label{inter function}
\end{equation}
Here, we have explicitly shown the dependence of $\eta$ on
${\mathbf{\Gamma}}$. Also, when we write ${\mathbf{\Gamma}}' \geq
{\mathbf{\Gamma}}$, we mean that $\Gamma_k'\geq \Gamma_k$ for
$k=1,\cdots,K$. Now, recall that ${\Gamma_k} =\frac{p_k
h_k}{\sigma^2}$. Hence, based on \eqref{inter function}, the UPC
algorithm can be expressed as
\begin{equation}\label{eqUPC}
{\mathbf{\Gamma}}(n+1)={\mathbf{I}}({\mathbf{\Gamma}}(n)) .
\end{equation}

\begin{proposition} For the matched filter, the decorrelator, and the MMSE detector,
if there exists a $\hat{{\mathbf{\Gamma}}}$ such that
$\hat{{\mathbf{\Gamma}}} \geq
{\mathbf{I}}(\hat{{\mathbf{\Gamma}}})$, then for every initial
vector ${\mathbf{\Gamma}}(0)$, the sequence
${\mathbf{\Gamma}}(n+1)={\mathbf{I}}({\mathbf{\Gamma}}(n))$
converges to the unique fixed point solution of
${\mathbf{\Gamma^*}}={\mathbf{I}}({\mathbf{\Gamma^*}})$.
Furthermore, ${\mathbf{\Gamma^*}} \leq  \hat{{\mathbf{\Gamma}}}$
for all $\hat{{\mathbf{\Gamma}}} \geq
{\mathbf{I}}(\hat{{\mathbf{\Gamma}}})$.
\end{proposition}\vspace{-0.2cm}
\begin{thmproof}{Proof:}
The condition that there exists a $\hat{{\mathbf{\Gamma}}} \geq
{\mathbf{I}}(\hat{{\mathbf{\Gamma}}})$ states that a feasible SNR
vector exists for achieving $\gamma^*$. To prove the proposition,
it is sufficient to show that ${\mathbf{I}}({\mathbf{\Gamma}})$ is
a standard interference function (see \cite{Yates95}),  i.e., for
all ${\mathbf{\Gamma}} \geq {\mathbf{0}}$, the following three
properties are satisfied.

1) Positivity: ${\mathbf{I}}({\mathbf{\Gamma}})>{\mathbf{0}}$; \
2) Monotonicity: If ${\mathbf{\Gamma'}}\geq{\mathbf{\Gamma}}$,
then
${\mathbf{I}}({\mathbf{\Gamma'}})\geq{\mathbf{I}}({\mathbf{\Gamma}})$;
\ 3) Scalability: For all $\theta>1$, $\theta
{\mathbf{I}}({\mathbf{\Gamma}})>{\mathbf{I}}({\mathbf{\theta
\Gamma}})$.

The dependence of the multiuser efficiency on $k$ is due to the
fact that different users may have different receivers. However,
we can assume, without loss of generality, that all users have the
same receiver type (and hence the same multiuser efficiency).
Therefore, to prove the proposition, it suffices to show that for
each receiver type, the three properties (i.e., positivity,
monotonicity, and scalability) are satisfied for
${\hat{I}({\mathbf{\Gamma}})=\frac{1}{\eta ({\mathbf{\Gamma}})}}$.

Positivity of $\hat{I}({\mathbf{\Gamma}})$ is trivial by
\eqref{eqA} for all receivers since $\eta \in [0,1]$.

For the matched filter, the multiuser efficiency is given by
\eqref{MF eta}, i.e., $\eta= \frac{1}{1+\alpha
{\mathbb{E}}\{\Gamma\}}$. Now, if
${\mathbf{\Gamma'}}\geq{\mathbf{\Gamma}}$, then
${\mathbb{E}}\{\Gamma'\}\geq{\mathbb{E}}\{\Gamma\}$. Therefore,
$\hat{I}({\mathbf{\Gamma'}})\geq \hat{I}({\mathbf{\Gamma}})$. To
prove the third property, note that, for $\theta>1$,
$\hat{I}(\theta {\mathbf{\Gamma}})=1+\alpha {\mathbb{E}}\{\theta
\Gamma\} < \theta +\alpha \theta {\mathbb{E}}\{\Gamma\}= \theta
\hat{I}({\mathbf{\Gamma}})$.

For the decorrelating detector, the multiuser efficiency is given
by \eqref{DE eta}, i.e., $\eta= 1-\alpha$ for $\alpha<1$. Since in
this case, $\eta$ is independent of ${\mathbf{\Gamma}}$, proving
properties 2 and 3 is straightforward.

For the MMSE detector, the multiuser efficiency is the solution to
\eqref{MMSE eta}, or equivalently, the solution of $\eta +\alpha
{\mathbb{E}}\left\{ \frac{\eta\Gamma}{1+\eta\Gamma}\right\}=1$.
Note that the left-hand side increases if both $\eta$ and $\Gamma$
increase. Thus if ${\mathbf{\Gamma'}}\geq{\mathbf{\Gamma}}$, we
must have $\eta({\mathbf{\Gamma'}})\leq \eta({\mathbf{\Gamma}})$
to maintain the equality. Hence, ${\hat{I}({\mathbf{\Gamma'}})\geq
\hat{I}({\mathbf{\Gamma}})}$. To prove the third property, let us
define $\eta'=\eta({\mathbf{\theta \Gamma}})$ and $\eta''=\theta
\eta'$, where $\theta>1$. Therefore, we have ${\eta' +\alpha
{\mathbb{E}}\left\{ \frac{1}{\frac{1}{\eta' \theta
\Gamma}+1}\right\}=1}$, or equivalently, $\eta'' + \alpha \theta
{\mathbb{E}}\left\{ \frac{1}{\frac{1}{\eta''
\Gamma}+1}\right\}=\theta$. Showing ${\theta
\hat{I}({\mathbf{\Gamma}}) > \hat{I}(\theta {\mathbf{\Gamma}})}$
is equivalent to showing $\eta'' > \eta$. Since ${\eta'' +\alpha
{\mathbb{E}}\left\{ \frac{1}{ \frac{1}{ \eta'' \Gamma}+1}\right\}
=1 + (1-\frac{1}{\theta})\eta'' > 1}$ and ${\eta +\alpha
{\mathbb{E}}\left\{ \frac{1}{ \frac{1}{\eta
\Gamma}+1}\right\}=1}$, and because ${\eta +\alpha
{\mathbb{E}}\left\{ \frac{1}{ \frac{1}{\eta \Gamma}+1}\right\}}$
is increasing in $\eta$, we must have $\eta''>\eta$. Therefore,
${\theta \hat{I}({\mathbf{\Gamma}}) > \hat{I}(\theta {\mathbf{\Gamma}})}$.\vspace{0.2cm}\\
This completes the proof.\hspace{11cm} \vspace{0.2cm} ${\Box}$
\end{thmproof}
In the following section, we study the performance of the UPC
algorithm for finite-size systems and compare it with that of
existing power control algorithms. A more detailed study of the
UPC algorithm can be found in \cite{MeshkatiTwireless}.

\section{Performance Evaluation and Discussion} \label{performance}

The existing SIR-based power control algorithms such as the ones
proposed in \cite{Foschini93} and \cite{Ulukus97}, update the
transmit powers of the users according to\vspace{-0.1cm}
\begin{equation}\label{eq99}
    p_k(n+1)=\frac {\gamma^*}{\gamma_k (n)} p_k (n) ,
\end{equation}
where $\gamma_k$ is the output SIR of user $k$. For the matched
filter, the decorrelator, and the MMSE detector, $\gamma_k$ is
expressed as
\begin{eqnarray}
    \gamma_k^{MF} &=& \frac{p_k h_k } {\sigma^2 + \sum_{j\neq k} p_j h_j ({\mathbf{s}}_k^T
    {\mathbf{s}}_j )^2 }\ ,\label{eq100}\\
    \gamma_k^{DE} &=& \frac{p_k h_k } {\sigma^2 \left[({\mathbf{S}}^T
    {\mathbf{S}})^{-1} \right]_{kk} }\ , \label{eq101}\\
    \gamma_k^{MMSE} &=& p_k h_k ({\mathbf{s}}_k^T {\mathbf{A}}_k^{-1}
    {\mathbf{s}}_k)\ \label{eq102},
\end{eqnarray}
where ${\mathbf{S}}=\left[{\mathbf{s}}_1 , {\mathbf{s}}_2 ,\cdots,
{\mathbf{s}}_K\right]$, ${\mathbf{A}}_k= \sum_{j\neq k} p_j h_j
{\mathbf{s}}_j {\mathbf{s}}_j^T + \sigma^2 \mathbf{I}$, and
$\left[({\mathbf{S}}^T {\mathbf{S}})^{-1} \right]_{kk} $ is the
$(k,k)$ entry of the matrix $({\mathbf{S}}^T{\mathbf{S}})^{-1}$.

The SIR-based power control algorithm in \eqref{eq99} cannot be
easily applied to the optimal multiuser receivers since finding
the output SIR for these receivers is not straightforward. In
addition, since the expressions for the output SIR are all
dependent on the spreading sequences of the users, in systems with
long spreading sequences, the SIR-based algorithm in \eqref{eq99}
has to continuously adjust the users' transmit powers as the
spreading sequences change from symbol to symbol even if the
channel gains stay unchanged. The UPC algorithm, on the other
hand, is a large-system approach and is, hence, independent of the
users' spreading sequences. Therefore, after convergence, the
users' transmit powers need not be updated as long as the channel
gains stay the same. Obviously, the true SIR \emph{does} depend on
the spreading sequences (as shown in
\eqref{eq100}--\eqref{eq102}). A question of interest is: if we
use the UPC algorithm, how close will the resulting SIRs be to the
target SIR? To answer this question, we focus on the decorrelating
and MMSE detectors.

\subsection{Decorrelating Detector}

We proved via Proposition 1 that the UPC algorithm converges to
the fixed point solution of
${\mathbf{\Gamma^*}}={\mathbf{I}}({\mathbf{\Gamma^*}})$. For the
decorrelating detector with $\alpha<1$, the multiuser efficiency
is given by $\eta^{DE}=1-\alpha$. As a result, we have
$\Gamma_k^*= \frac{\gamma^*}{1-\alpha}$, for $k=1,\cdots,K$.
Therefore, based on \eqref{eq101}, the true output SIR for the
decorrelating detector, in this case, is given by
\begin{equation}\label{eq104}
  \gamma_k= \left(\frac{\gamma^* }{1-\alpha} \right) \frac{1} { \left[({\mathbf{S}}^T
    {\mathbf{S}})^{-1} \right]_{kk} } \ .
\end{equation}
It can be shown that in systems with large processing gains, the
distribution of $\frac{1} {
\left[({\mathbf{S}}^T{\mathbf{S}})^{-1} \right]_{kk}}$ can be
approximated by a beta distribution with parameters $(N-K+1, K-1)$
\cite{Muller}. As a result, for the decorrelator, the probability
density function (PDF) of $\gamma_k$ is given approximately by
\begin{equation}\label{eq105}
     f_{\gamma_{DE}}(z) =
    \left(\frac{1}{\Gamma_{DE}^*}\right)^{N-1} \frac{
    z^{N-K}(\Gamma_{DE}^*-z)^{K-2}}{B(N-K+1, K-1)} \ \
    \textrm{with} \ \ z\leq \Gamma_{DE}^* \  ,
\end{equation}
where $B(a,b)=\int_0^1 t^{a-1} (1-t)^{b-1} \textrm{d}t$ and
$\Gamma_{DE}^*= \frac{\gamma^*}{1-\alpha}$. Therefore, as the
spreading sequences change from symbol to symbol, the probability
that $\gamma_k$ stays within $\Delta$~dB of $\gamma^*$ is given by
\begin{equation}\label{eq106}
    P_{\Delta, DE}\equiv \textrm{Pr}\left\{ |\gamma_{DE} (dB) -
    \gamma^* (dB) | \leq \Delta\right\} =
    \int_{\gamma_L}^{\gamma_H} f_{\gamma_{DE}}(z)\textrm{d}z \
    ,
\end{equation}
where $\gamma_L=10^{-\frac{\Delta}{10}}\gamma^*$ and
$\gamma_H=10^{\frac{\Delta}{10}}\gamma^*$.

\subsection{MMSE Detector}

If all users have the same target SIR, $\gamma^*$, the
steady-state SNRs will be identical after the UPC algorithm
converges (i.e., $\Gamma_1^*=\cdots=\Gamma_K^*=\Gamma^*$). The
multiuser efficiency in this case will be given by\vspace{-0.15cm}
\begin{equation}\label{eq107}
    \eta_{MMSE}=\frac{1-\alpha}{2}-\frac{1}{2\Gamma_{MMSE}^*} +
    \sqrt{\left(\frac{1-\alpha}{2}\right)^2 +
    \frac{1+\alpha}{2\Gamma_{MMSE}^*} +\left(\frac{1}{2\Gamma_{MMSE}^*}\right)^2}  \
    ,
\end{equation}
with $\Gamma_{MMSE}^*=
\frac{\gamma^*}{1-\alpha\frac{\gamma^*}{1+\gamma^*}}$ assuming
that $\alpha<1+\frac{1}{\gamma^*}$. It can be shown that for the
MMSE detector, the fluctuation of the true SIR around $\gamma^*$
is approximately Gaussian with variance $\frac{c}{N}$
\cite{TseZeitouni, KimHonig}, where
$c=\frac{2{\gamma^*}^2}{1-\alpha\left(\frac{\gamma^*}{1+\gamma^*}\right)^2}$,
i.e., $\gamma_{MMSE}\sim{\mathcal{N}}(\gamma^*, \frac{c}{N})$.
Hence, the probability that $\gamma_k$ stays within $\Delta$~dB of
$\gamma^*$ is given approximately by\vspace{-0.15cm}
\begin{equation}\label{eq109}
    P_{\Delta,MMSE} \equiv \textrm{Pr}\left\{ |\gamma_{MMSE}
\textrm{(dB)} -
    \gamma^* \textrm{(dB)} | \leq \Delta\right\}
    \simeq \Phi\left(\sqrt{\frac{N}{c}}~(\gamma_H-\gamma^*)\right)-
   \Phi\left(\sqrt{\frac{N}{c}}~(\gamma_L-\gamma^*)\right) \ ,
\end{equation}
where $\Phi(\cdot)$ is the cumulative distribution function of the
standard normal distribution.

It is seen that for both the decorrelator and the MMSE detector,
the variance of fluctuations of SIR decreases as $1/N$. In the
following section, we demonstrate the performance of the UPC
algorithm using simulations and also investigate the accuracy of
the theoretical approximations discussed above.

\section{Numerical Results}\label{simulation}

We consider the uplink of a DS-CDMA system with $K$ users and
processing gain $N$, with random (long) spreading sequences. The
background noise  power, $\sigma^2$, is assumed to be ${1.6\times
10^{-14}}$Watts and the target SIR, $\gamma^*$, is equal to 6.4
(i.e., $8.1$~dB). Our choice for the target SIR comes from the
results in \cite{Meshkati_Tcomm}.

\begin{figure}
\begin{center}
\leavevmode \hbox{\epsfysize=9cm \epsfxsize=10cm
\epsffile{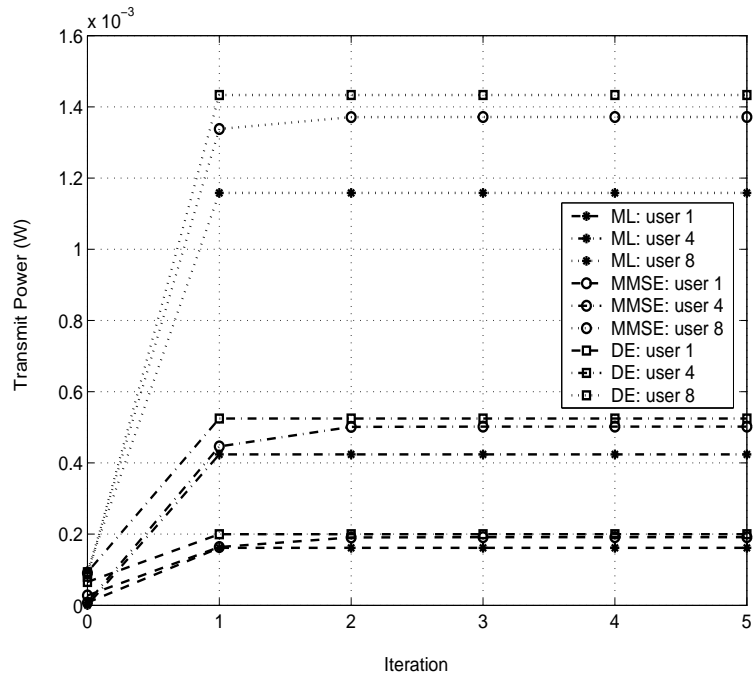}}
\end{center}
\vspace{-0.5cm}\caption{\small Users' transmit powers for the ML,
MMSE, and decorrelating detectors, using the UPC algorithm ($N=32$
and $K=8$).} \label{figsim1}
\end{figure}

We first demonstrate the convergence of the UPC algorithm by
considering a system with 8 users and spreading factor 32 (i.e.,
$K=8$ and $N=32$). The channel gain for user $k$ is given by
$h_k={0.1}{d_k^{-4}}$ where $d_k$ is the distance of user $k$ from
the uplink receiver (e.g., base station) and is assumed to be
given by $d_k=100+10k$ in meters. We implement the UPC algorithm
for the decorrelator and the MMSE detector as well as the maximum
likelihood (ML) detector (which is equivalent to the jointly
optimal detector). In Fig.~\ref{figsim1}, we show the transmit
powers for users 1, 4, and 8 at the end of each iteration. It is
seen that for all three receiver types, the UPC algorithm
converges very quickly to steady-state values. The results are
similar when the initial power values and/or $K$ and $N$ are
changed. It is also observed that the steady-state transmit powers
for the decorrelator and the MMSE detector are close to those of
the ML detector (in this case, the difference is less than 22\%).
This means that in terms of energy efficiency\footnote{We define
energy efficiency as the utility (in bits per joule) achieved by
the users in the network at the Nash equilibrium. The utility
function of a user is the ratio of the user's throughput to its
transmit power (see \cite{Meshkati_Tcomm}).}, the decorrelator and
the MMSE detector are almost as good as the ML detector.

\begin{figure}
\begin{center}
\leavevmode \hbox{\epsfysize=9cm \epsfxsize=10cm
\epsffile{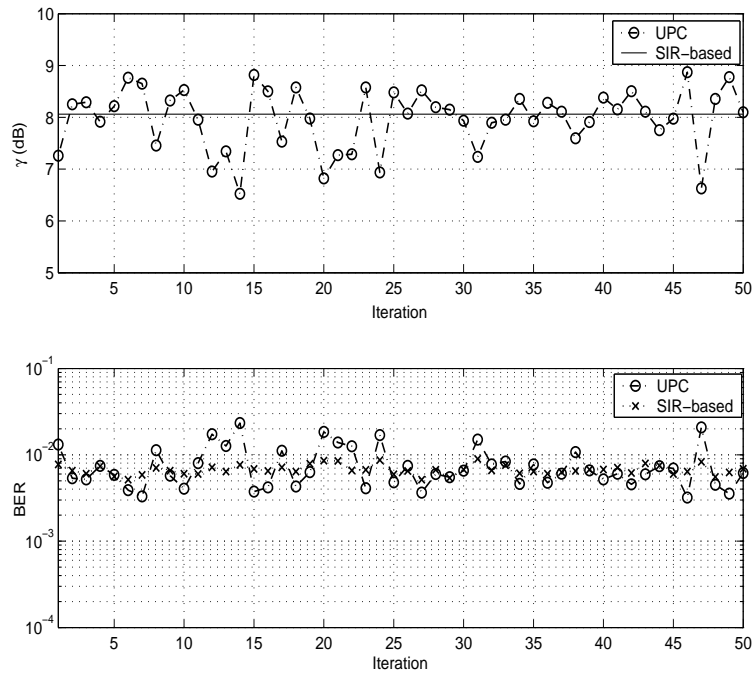}}
\end{center}
\caption{ User 1 output SIR and BER for the UPC algorithm and
SIR-based algorithm with the MMSE detector ($N=32$ and $K=8$).}
\label{MMSE-gamma-ber}
\end{figure}

We now compare the performance of the UPC algorithm with that of
the SIR-based algorithm of \eqref{eq99}. Fig.~\ref{MMSE-gamma-ber}
shows the SIR and bit-error-rate (BER) of user~1 for the UPC and
SIR-based algorithms for the MMSE detector. It is seen that the
SIR-based algorithm achieves the target SIR, $\gamma^*$, at all
time whereas the output SIR for the UPC algorithm fluctuates
around the target SIR as the spreading sequences change. It should
be noted that the BER values in this plot fluctuate around
$Q(\sqrt{\gamma^*})=1-\Phi(\sqrt{\gamma^*})$=0.006 which is the
BER corresponding to $\gamma^*=6.4$ (assuming additive Gaussian
noise/interference).
\begin{figure}
\begin{center}
\leavevmode \hbox{\epsfysize=11.5cm \epsfxsize=11cm
\epsffile{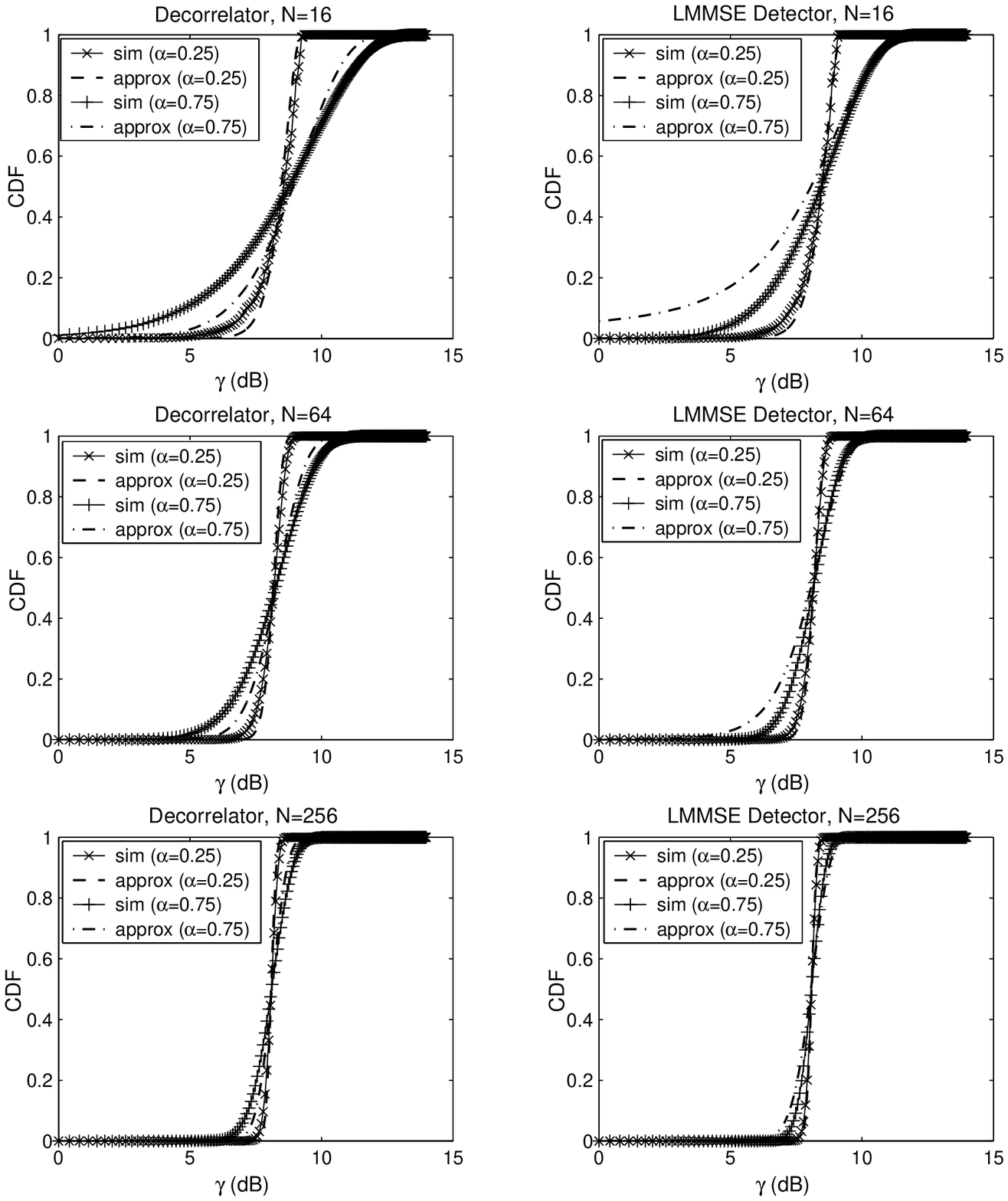}}
\end{center}
\vspace{-0.5cm}\caption{\small Cumulative density functions (CDFs)
of $\gamma$ for the decorrelating and MMSE detectors.}
\label{figsim4}
\end{figure}

To evaluate the accuracy of the theoretical approximations given
in Section~\ref{performance}, we have plotted the cumulative
probability distribution functions (CDFs) of $\gamma$ for the
decorrelating and MMSE detectors for different processing gains
with both low and high system loads in Fig.~\ref{figsim4}. In this
figure, we have plotted the CDFs obtained from simulation (based
on 100,000 realizations) as well as those predicted by the
theoretical approximations. It is seen from the figure that the
theoretical approximations become more accurate as the processing
gain increases. Also, in general, the approximations are more
accurate when the system load is low. This figure suggests that
the UPC algorithm is more useful when the processing gain is high
and/or the system load is low.

To quantify the discrepancies between the simulation results and
the theoretical approximations, we have computed $P_{\Delta, DE}$
and $P_{\Delta, MMSE}$ using the CDFs obtained from simulation as
well as those predicted by theory (see \eqref{eq106} and
\eqref{eq109}). Table \ref{table1} shows the results for different
processing gains and system loads for $\Delta=1$~dB. The numbers
in the table represent the probability that $\gamma$ is within
1~dB of $\gamma^*$. The probabilities obtained by simulation
suggest that the UPC algorithm performs better for the MMSE
detector than for the decorrelator. It is also seen from the table
that when the processing gain is small, the fluctuation in the
output SIR is considerable, especially when the system load in
high. The performance improves as the processing gain increases.
For example, for the MMSE detector, when $N=256$ and
$\alpha=0.75$, the SIR stays within 1~dB of the target SIR $98\%$
of the time. It is also observed that the theoretical
approximations are optimistic for the decorrelator and very
pessimistic for the MMSE detector.
\begin{table}
\begin{center}
\caption{\small Summary of results for the decorrelating and MMSE
detectors}\label{table1}\vspace{0.1cm} {\footnotesize{
\begin{tabular}{|c |c |c |c |c |c |c |c |c|}
  \hline
   N    & $P_{\textrm{1dB}, DE}^{Sim} $  & $P_{\textrm{1dB}, DE}^{Approx}$ & $P_{\textrm{1dB},
   DE}^{Sim} $ & $P_{\textrm{1dB}, DE}^{Approx}$   & $P_{\textrm{1dB}, MMSE}^{Sim}$
   & $P_{\textrm{1dB}, MMSE}^{Approx} $ & $P_{\textrm{1dB},
   MMSE}^{Sim} $ & $P_{\textrm{1dB}, MMSE}^{Approx}$  \\
    & $\alpha=0.25$ & $\alpha=0.25$ & $\alpha=0.75$ &
    $\alpha=0.75$& $\alpha=0.25$ & $\alpha=0.25$ & $\alpha=0.75$ &
    $\alpha=0.75$\\
  \hline \hline
   16 & 0.77 & 0.87 & 0.28 & 0.19 & 0.93 & 0.46 & 0.41 & 0.33 \\
   64 & 0.98 & 1.0 & 0.54 & 0.64  & 0.99 & 0.76 & 0.74 & 0.61 \\
   256 & 1.0 & 1.0 & 0.87 & 0.96  & 1.0  & 0.98 & 0.98 & 0.91 \\
   \hline
\end{tabular}}}
\end{center}
\end{table}

\section{Conclusions}\label{conclusions}

A unified power control (UPC) algorithm which is applicable to a
large family of detectors including many of the most widely
studied multiuser detectors has recently been proposed. In this
work, we have studied the convergence and performance of the UPC
algorithm. In particular, we have proved the convergence of the
algorithm for the matched filter, the decorrelator, and the
(linear) MMSE detector. In addition, the performance of the
algorithm in finite-size systems with long spreading sequences has
been studied and compared with that of the existing power control
schemes. Theoretical approximations for predicting the performance
of the UPC algorithm have been presented and their accuracies have
been investigated using simulations. We have shown that in systems
where achieving the target SIR is crucial to the performance, the
UPC algorithm is useful primarily when the processing gain is
large and/or the system load is small.


{\footnotesize{
 }}

\end{document}